# Bias sputtering of granular $L1_0$-FePt films with hexagonal boron nitride grain boundaries

Chengchao Xu[1, 2, *], B.S.D.Ch.S Varaprasad[1, 2], David E. Laughlin[1, 2, 3], and Jian-Gang (Jimmy) Zhu[1, 2, 3].

1) Data Storage Systems Center, Carnegie Mellon University, Pittsburgh, Pennsylvania 15213, USA
2) Electrical and Computer Engineering Department, Carnegie Mellon University, Pittsburgh, Pennsylvania 15213, USA
3) Materials Science and Engineering Department, Carnegie Mellon University, Pittsburgh, Pennsylvania 15213, USA

* Correspondence and requests for materials should be addressed to C.X. (email: chengchx@andrew.cmu.edu)

## Abstract

In this paper, we present an experimental study of $L1_0$-FePt granular films with crystalline/amorphous boron nitride (BN) grain boundary materials for heat assisted magnetic recording (HAMR). It is found that an adequate RF substrate bias yields the formation of hexagonal boron nitride (*h*-BN) nanosheets in grain boundaries, facilitating the columnar growth of FePt grains during sputtering at high temperatures. The *h*-BN monolayers conform to the side surfaces of columnar FePt grains, completely encircling individual FePt grains. The resulting core-shell FePt/*h*-BN nanostructures appear to be highly promising for HAMR application. The high thermal stability of *h*-BN grain boundaries allows the deposition temperature to be as high as 800°C such that high order parameters of FePt $L1_0$ phase have been obtained. For the fabricated FePt/*h*-BN thin film, excellent granular microstructure with FePt grains of 6.5nm in diameter and 11.5nm in height has been achieved along with good magnetic hysteresis properties.

# Introduction

For heat-assisted magnetic recording (HAMR), $L1_0$-FePt granular thin films have been the choice of recording media mainly due to the material's ultra-high magneto-crystalline anisotropy and its characteristic temperature dependence near the Curie point. Over the past two decades, extensive research and development efforts have been devoted to improving the microstructure with relatively thermal-insulating grain boundary materials (GBM) to fabricate columnar-shaped tall (height > 10nm), small (diameter < 8nm) and $L1_0$-ordered (order parameter > 0.75) FePt grains[1,2]. The formation of the $L1_0$ phase of FePt requires elevated substrate temperatures during deposition, which presents a great challenge for achieving the desired microstructure. Searching for an ideal GBM that enables the desired microstructure at a relatively high deposition temperature, commonly above 600°C, has been the constant focus of research in this area. A list of GBM, particularly carbon, various oxides, and boron nitride, has been given tremendous attention over the past.

Although amorphous carbon as the GBM enables well-separated grains, it fails to promote columnar growth[3–5]. The spherical shape of the FePt grains in the FePt-C film limits the grain height to be similar to the in-plane grain diameter. Attempts to produce taller grains always result in forming second layers of FePt grains that are not correctly oriented and poorly ordered. Oxides like $TiO_2$[6,7], $SiO_2$[8–10], and $TaO_x$[11] have been found to facilitate columnar growth; however, their deficiency in either microstructure or $L1_0$ ordering has prevented any success from actual adoption. For example, silicon oxide typically enables columnar growth but often fails to encircle FePt grains, causing a good percentage of neighboring FePt grains to become laterally connected[8]. It has been suggested that the actual melting temperature of the sputtered $SiO_2$ GBM might be substantially lower than that of the bulk form[10]. In order to inhibit the lateral growth of FePt grains at high process temperatures, a GBM with higher thermal stability is required. Amorphous boron nitride (*a*-BN) as a GBM has some advantages over carbon and $SiO_2$[12–14]. However, it was shown that adding *a*-BN appears to suppress the $L1_0$ ordering of FePt, likely due to the boron diffusion into FePt lattice, creating localized Fe-B bonding[15]. Several techniques have been tried to improve the ordering of FePt with *a*-BN, including adding $N_2$ to the sputtering atmosphere[16], mixing with other GBM[17,18], and applying DC substrate bias[19]. Nonetheless, the outcomes are far from satisfaction.

What if we can replace *a*-BN with crystalline BN for the grain boundaries? The crystalline hexagonal boron nitride (*h*-BN) has a staggered AB stacking of hexagonal honeycomb monolayers. Adjacent monolayers have a fixed spacing determined by the interlayer Van de Waals forces[20]. The *h*-BN has excellent thermal stability and chemical inertness. All these attributes could be advantageous to serve as GBM for granular $L1_0$-FePt thin films. Here, we present a systematic experimental investigation of the fabrication of such FePt/*h*-BN granular thin film. X-ray diffraction (XRD), transmission electron microscopy (TEM) analysis and magnetic hysteresis measurements have been performed to evaluate the microstructure and properties. Concerns over lateral thermal conductivity, which is important for HAMR recording performance, will be addressed at the end of this paper.

**Results and Discussion**

Figure 1(a) shows the plane-view high resolution TEM image of a FePt/ *h*-BN sample of a thickness $t$=7nm and a BN concentration of 38% (referred to as Sample-1). The film sample was fabricated in the following steps: (1) a 0.5nm thick FePt was initially sputtered onto an 8nm thick MgO underlayer, (2) an approximate 1nm thick FePt-BN layer was then deposited using the co-sputtering technique with separate $Fe_{0.55}Pt_{0.45}$ and amorphous boron-nitride targets without bias on the substrate, and (3) continued co-sputtering of FePt and BN with RF bias applied on the sample substrate at a bias RF power of 4W. From Step (1) to Step (3), the substrate was heated and maintained at 650°C with a total deposition time of 6 minutes.

The image in Fig. 1(a) shows that the FePt grains are well separated by crystalline boron nitride monolayers conforming to grain boundary surfaces and encircling each grain. The FFT (fast Fourier transform) analysis (Fig. 1(b)) of this image reveals a diffraction ring over the range of 0.33 nm~0.35 nm, matching the *d*-spacing of *h*-BN (0002) planes ($d$ = 0.333 nm). In addition, the presence of the FePt {110} superlattice spots shown in the FFT pattern demonstrates the L1$_0$ ordering and the [002] texture of FePt grains in this film.

The cross-sectional view shown in Fig. 1(d) of a different sample, referred to as Sample-2, fabricated with a similar procedure as Sample-1 but with a lower BN volume fraction (detail shown in Fig. 2(a)), adds more details of the crystalline nanostructure for both FePt grains and boron nitride grain boundaries. The BN material deposited in Step (2) without substrate bias is amorphous. In this step, while growing with the *a*-BN boundary layer, the FePt grains are shown to grow with specific crystalline facets which match the equilibrium shape (Wulff construction) of the L1$_0$ FePt nanocrystal: a truncated octahedron made up of eight {111} planes and six {100} planes[21–23]. The illustration in Fig. 1(e) shows the view of the Wulff polyhedron along [$\bar{1}$10] direction (blue dashed line) matching the bottom of FePt grains, in agreement with the TEM image on the left. Due to the faceted growth in this step, the lateral sizes of the FePt grain increased with film thickness. Fig. 1(e) also shows that when the bias is applied on the substrate, *h*-BN nanosheets start to form, with the honeycomb monolayer growing parallel to the side surfaces of the FePt grains. The *h*-BN evidently stops the lateral growth of the FePt grains since the number of *h*-BN monolayers inside a grain boundary remains unchanged as the FePt grain and the *h*-BN grain boundary grow taller simultaneously. The TEM image in Fig. 1(d) clearly shows the constant grain boundary width through the film thickness as the number of *h*-BN monolayers in grain boundaries remains unchanged.

To reiterate, the formation of *h*-BN monolayers in the grain boundaries yields the columnar growth of the FePt grains with grain size and grain boundary gap largely unchanged, whereas the initial FePt/*a*-BN layer results in faceted FePt grain growth that leads to grain size increase with film thickness. Tuning the onset point of the bias application during film deposition enables the control of the FePt/*a*-BN layer thickness and, thereby, the FePt grain size.

With the understanding obtained, the deposition Step (3) is refined as shown in Fig. 2(a) to achieve a microstructure more suitable to HAMR application, that is, one with narrower grain boundaries. The produced sample is Sample-2, referred to in Fig. 1(d) above. In Step (3) boron nitride concentration was varied through the film thickness with an average volume fraction of 19%. Again, over the entire FePt/FePt-BN deposition process, the substrate temperature was maintained at 650°C. The plane-view TEM image of this sample is shown in Fig. 2(b). The well-defined granular microstructure shows that the majority of FePt grains are completely encircled by the BN grain boundaries, with sparsely scattered exceptions where some adjacent FePt grains are laterally connected. Statistical distributions of grain sizes and center-to-center pitch distances based on TEM images are shown in Fig. 2(c) and 2(d) respectively. The results show an average grain size of 6.7 nm and an average grain pitch of 8.2 nm. Consequently, the average grain boundary width is around 1.5nm as the number of *h*-BN monolayers within a grain boundary is around four on average, matching what is shown in cross-section TEM images. Figure 2(e) shows a typical cross-sectional image of the same sample (high-resolution image in Fig. 1(d) and supplementary Fig. S1). Most grains are columnar with an average grain height of 11.5 nm. Note that even when some grains are tilted from the film normal, the grain boundary gap width appears to remain the same through the thickness, with the side surfaces of the two adjacent tilted grains growing in parallel. Though most FePt grains have developed tall and columnar shapes with an aspect ratio $h/D = 1.73$, imperfect grains can be clearly seen, which will be characterized later in the paper.

The formation of *h*-BN is the key factor for achieving good granular microstructures. Since the film is formed at relatively high substrate temperatures during the deposition, good atomic ordering of the FePt L1$_0$ phase can be readily achieved. Figure 3(b) shows the θ–2θ X-ray diffraction (XRD) pattern of Sample-2. Note that no (111) peak can be observed, indicating good [001] texture for FePt grains. The integrated intensity ratio of the L1$_0$-ordered superlattice peak (001) to the fundamental peak (002) ($I_{001}/I_{002}$) is about 2.53, and the calculated order parameter[24] is $S = 0.782$. A barely visible FePt (200) peak located at the left shoulder of the relatively sharp L1$_0$ (002) peak indicates a very small percentage of in-plane oriented grains. The FWHM's of L1$_0$ (001) and (002) peaks in the measured rocking curves are both around 9.5°. Figure 3(a) shows the perpendicular and in-plane magnetic hysteresis loops, measured at room temperature, with a perpendicular coercivity of 37.7 kOe and an in-plane coercivity of 5.38 kOe. The opening of the in-plane curve likely results from the combination of in-plane ordered L1$_0$ FePt grains and the relatively broad dispersion of FePt c-axis orientation.

Plane-view STEM images in high-angle annular dark field mode (HAADF) have been taken for film samples with the deposition processes stopped at different stages for the film stack shown in Fig. 2(a). Figures 4(a), 4(b), and 4(c) show the images for three different samples, referred to as Sample-2α ($t$ =3.5nm), Sample-2β ($t$ =7.5nm) and Sample-2 ($t$ =11.5nm), respectively. The false color in the images represents the lateral size of FePt grains: larger grains are colored more to the yellow and smaller grains are colored more to the cyan. The image shown in Fig. 4(a), Sample-2α, shows the plan view of FePt grains mainly with *a*-BN grain boundaries that are deposited without substrate bias. Most grains have square or rectangular shapes with edges parallel to <110> directions (see Supplementary Fig. S3(a)), which agrees with the top-down view (along [001] direction) of the Wulff polyhedron of FePt and is consistent with the faceted growth analysis provided earlier. There also exists a significant percentage of very small grains as the bimodal distribution of grain sizes is evident at this early growth stage. At the film thickness of 7.5 nm, as shown in the image of Fig. 4(b), not only did the lateral grain size become larger compared to Sample-2α, but also the bimodal distribution shown in Fig. 4(a) has evolved to a single peak with the disappearance of the smaller grain size peak. Since the substrate bias had already started by the moment of 3.5nm film thickness, *h*-BN should have been forming in grain boundaries during the film growth between Sample-2α and Sample-2β. Therefore, it is reasonable to think that adjacent grains with grain boundary separation less than the space needed to fit in at least one *h*-BN monolayer are likely to merge, especially when the two grains are nucleated on the same MgO grain of the underlayer. This is one possible mechanism for the disappearance of the bimodal distribution. We will discuss other possible mechanisms later.

Comparing the two images shown in Fig. 4(b) and Fig. 4(c), the grain size and its distribution film are basically unchanged as the film grows from 7.5nm to 11.5nm in thickness. This is consistent with the fact that FePt grains grow largely columnar as facilitated by the *h*-BN monolayers in the grain boundaries. However, in both images, laterally connected FePt grains can be seen, whereas they are not present at the 3.5nm case (Fig. 4(a)). These connected grains will be discussed in detail when we get to Fig. 5.

Figure 4(d) illustrates our hypothesis for the reason that the application of substrate bias facilitates the formation of *h*-BN in the grain boundary. We believe that the re-sputtering caused by substrate bias is an important factor. Without substrate bias, deposited BN in the grain boundary is amorphous. The applied RF bias induces energetic ion bombardments that preferentially knock off weakly bonded B/N species, whereas strongly bonded BN can survive. Thus, the *h*-BN phase is much more likely to survive due to the strong covalent bonds between boron and nitrogen atoms within the (0002) hexagonal monolayers.

The illustration in Fig. 5(a) provides a detailed characterization of normal columnar growth for 90% of the grains; Figs. 5(b) and 5(c) illustrate two major types of defective growth for the other 10% of the grains. The 3.5nm thick case (upper half parts) reflects the bimodal nucleation. Fig. 5(b) illustrates (i) How a small grain nucleated adjacent to a much larger grain can grow towards the larger grain and eventually merges with it to become a single grain; (ii) a small grain nucleated in between two relatively larger grains can be shadowed during the film growth, providing the chance for boron nitride material to grow over the grain, a mechanism for forming a wider *h*-BN grain boundary; (iii) lateral coalesce of a large grain with a small grain close by in the early stage of the growth. Note that since the MgO grain size of the underlayer is much larger than that of FePt grains, the smaller and larger grains in both cases (i) and (iii) are more likely to be on top of the same MgO grain in which case the lattices of the two FePt grains would match to form a single crystal. Fig. 5(c) illustrates a different kind of defective growth that happens at relatively later stage of the film growth. In the case that an *h*-BN grain boundary grows slower than that of the two FePt grains at either side, the two grains can growth laterally, starting to shield the arrival of the B/N species and eventually close up the grain boundary gap from both sides. Several top-connected grains can be seen in the TEM image shown in Fig. 4(c), though they are sparsely distributed. High resolution cross-sectional TEM images in Figs. 5(d), 5(e) and Supplementary

Figs. S3(d)-(e) show clear evidence for the illustrations in Figs. 5(a), 5(b) and 5(c) and the accompanied discussion.

To understand why the *h*-BN nanosheets form conformal to the FePt grain side surfaces, a modeled experiment was carried out to study the formation of *h*-BN on a metal surface under the same exact sputtering condition. In this experiment, a 13-nm-thick boron nitride film was sputtered on a curved Cr surface with the substrate heated to 650°C and an RF bias power of 4W applied. Figure 6(a) shows a cross-sectional HRTEM image of the film as the lattice fringes of boron nitride basal planes can be clearly observed. The electron energy loss spectrum (EELS), Fig. 6(b), of the same sample shows four peaks in the Boron-K edge: one $\pi^*$ peak at 191.8 eV and three $\sigma^*$ peaks at 198.8 eV, 204.2 eV, 215.2 eV respectively, matching the characteristic peaks of the standard *h*-BN[25,26].

Going back to Fig. 6(a), for the initial 6~7 monolayers of *h*-BN at the Cr interface, the basal planes grow parallel to the curved Cr surface, regardless of the curvature of the surface. Namely, on either convex or concave surface locations, the *h*-BN monolayers are all conformal to the metal surface. This phenomenon thus seems not to have any correlation with Cr lattice orientation, nor do we see any evidence of epitaxy in our electron diffraction patterns. The observation here is in good agreement with the achieved core-shell FePt/*h*-BN microstructure, although we have a different metal, Cr, instead of FePt. At certain growth conditions, the formation of *h*-BN monolayers conforming to metal surface appear quite characteristic, at least for certain metals[27], including FePt and Cr shown in this study.

In addition, it is interesting to note that away from the interface, the basal planes of *h*-BN gradually turn to be oriented in the direction normal to the interface, a 90° orientation change from the interface. The perpendicular *h*-BN basal planes were proven to be the energy-favored orientation since the biaxial compressive stress builds up as the film grows thicker[28]. In a way, the observation confirms that the *h*-BN monolayers conforming to the metal surfaces are indeed an interfacial phenomenon. The consistent formation of *h*-BN monolayers conformal to the metal surface in both the modeled experiment and granular FePt/h-BN thin films is likely to be governed by the same underlying physics. Based on the facts observed in these experiments, along with other studies reported in the literature, we arrived at the following explanation: (1) The combined effect of substrate bias and high temperature makes *h*-BN the only surviving phase on the surface of the metal due to the strong covalent bonds within the (0002) monolayers, whereas the relatively weaker bonded *a*-BN would be desorbed away; (2) The RF bias induces more ion bombardments on the growing surface, gradually raising the in-plane compressive stress due to ion peening effect[19,29]. It may be a contributing factor for the perpendicular orientation of h-BN layers in both Fig. 1 and Fig. 6(a). (3) Previous studies have found that for *h*-BN honeycomb monolayer on various metal surfaces[30], the electronic structure of the *h*-BN monolayer remains largely unchanged due to strong intralayer bonding. It is also known that *h*-BN monolayers exhibit localized charge centers[29] (these charge centers are responsible for shifted matching for multilayer stacking and give rise to the van der Waals interactions between the layers). The conductive metal surface could behave as a charge "mirror", creating imaging charges of the charge centers in the monolayer, creating attractive forces causing the monolayer basal planes to be parallel to the metal surface.

## Summary and Remarks

A systematic experimental study was conducted to fabricate granular $L1_0$-FePt thin film media with boron nitride GBM. The film media were deposited using co-sputtering technique with separate targets of FePt and *a*-BN. Heating the substrate above 650°C, the initial deposition creates

FePt grains with faceted growth surrounded with *a*-BN grain boundaries, and FePt grain size increases with increasing film thickness. At an appropriate point, an RF bias of adequate power was applied to substrate. The substrate RF bias yields the formation of *h*-BN monolayers inside grain boundaries with their basal planes parallel to the side surfaces of FePt grains during subsequent film deposition. The number of *h*-BN monolayers mostly remained constant in a grain boundary through the deposition, facilitating a columnar growth for most of the FePt grains in the film sample. The study shown here indicates that it is possible to grow very tall FePt grains without increasing their lateral size while keeping good granular microstructure as long as the vertical growth rate of FePt grains and the vertical growth rate of *h*-BN inside grain boundaries can always be matched during film deposition. The rate matching should prevent the lateral connection between adjacent FePt grains. Another advantage is the high-temperature endurance of the granular microstructure. The films deposited at substrate temperatures as high as 800°C have similar microstructures as the ones shown here, which allows FePt grains to achieve high order parameters of the $L1_0$ phase.

For HAMR application, the thermal conductivity between adjacent FePt grains is naturally an important concern. Before any discussion, we need to note that the thermal conductivity in *h*-BN is highly anisotropic: the lateral thermal conductivity within each honeycomb monolayer is relatively high[31] ($\kappa_{//} > 200$ W/(m·K)), and the thermal conductivity perpendicular to the plane of the honeycomb monolayer is two orders of magnitude lower[32] ($\kappa_{\perp} \approx 2$ W/(m·K)). For the granular FePt/*h*-BN film fabricated here, thermal conductivity is likely to be dominated by the perpendicular thermal conductivity though it needs to be confirmed experimentally because of possible thermal conduction tangent to FePt grain boundaries. Nevertheless, experimental characterization of lateral thermal conductivity for the media achieved here is very much needed in future investigations.

## Methods

**Sample preparation**

In this study, all thin film samples were deposited on Si (001) substrates using an AJA sputtering system with base pressures of $2\times10^{-8}$ Torr or less. The underlayer consisted of a 4-nm amorphous Ta adhesion layer and a 30-nm Cr with a (002) texture to produce large grains and enhance the texture of subsequent films. As the seed layer for $L1_0$-FePt grains to form the chemical ordering and c-axis normal to the film plane, 8 nm of MgO was epitaxially deposited on Cr. The underlayer's growth parameters are selected to achieve a low grain boundary density, low surface roughness, and a strong MgO [002] texture (rocking curve <= 8°). The epitaxial relation is MgO (002) <002> // $L1_0$-FePt (002) <002>, and the lattice mismatch is around 10%.

The magnetic layers are composed of $L1_0$ FePt grains and oriented $h$-BN nanosheets, deposited by co-sputtering $Fe_{0.55}Pt_{0.45}$ and $a$-BN targets at 650°C and 5 mTorr of Ar in the chamber. The distance from substrate to targets was roughly 20 cm. A 3~4W RF bias was applied to the substrates of certain samples and the substrate voltage at steady state was ~15V. The deposition rate of $h$-BN is quite low (~0.007 nm/s) due to the substrate bias. All samples' recording layers were fabricated in a multi-step way, similar to what is illustrated in Fig. 2(a). The initial layer (designated as [L0] layer) was a 0.5-nm pure FePt nucleation layer. Then the co-sputtered FePt-BN layers always began with a [L1] layer deposited without substrate bias, followed by a [L2] layer containing the same volume fraction of BN with bias. It was discovered that when the substrate bias was applied to the entire magnetic layer, the resulting microstructure and ordering were poor. This delayed bias was intended to prevent the disruption of the nucleation stage and was shown to improve the final microstructure. For samples of different thicknesses, one to three sublayers with slightly lower BN volume fractions were then deposited. Essentially, the BN vol % decreases gradually from the bottom sublayer to the top. It has been demonstrated that this stack with graded volume fractions prevents GBM from capping the top of FePt grains, thereby inhibiting the formation of second layers in the FePt-SiO$_x$ system. It turns out to be effective in the FePt-BN system as well. We denote the sample with 38vol% BN and thickness of 7.5nm as Sample-1, whose film stack is FePt (0.5 nm) / FePt-40vol% $a$-BN (1 nm) / FePt-40vol% $h$-BN (6 nm). We denote the sample with 19vol% BN and thickness of 11.5nm as Sample-2, whose film stack is FePt (0.5 nm) / FePt-22vol% $a$-BN (1.2 nm) / FePt-22vol% $h$-BN (4.9 nm) / FePt-18vol% $h$-BN (2.6 nm) / FePt-16vol% $h$-BN (2.3 nm). The film stack of Sample-2 is also shown in Figure 2(a).

**Sample characterizations**

The degree of chemical ordering of FePt and crystalline texture of all the layers were examined using the traditional X-ray diffraction (XRD) technique with Cu Kα radiation. The superconducting quantum interface device vibrating sample magnetometer (SQUID-VSM) (Quantum Design MPMS3 system) with an applied magnetic field up to 7T was used to measure the magnetic properties at room temperature. The plane-view and cross-sectional transmission electron microscopy (TEM) imaging was used to assess the microstructure of the samples, including bright-field TEM (BF-TEM), high-resolution TEM (HR-TEM), scanning TEM-high angle annular dark field and bright field (STEM-HAADF & STEM-BF) techniques. For accuracy and consistency, the STEM-HAADF plan-view images and image processing software MIPAR were used to analyze the grain size and grain center-to-center pitch distances. For each grain, the pitch distance analysis counted its six closest neighbors. The in plane- STEM-HADDF images of the grains were colored based on the in-plane grain diameters, where larger ones were colored cyan and smaller ones were colored yellow.

## Data Availability

All data gathered and/or analyzed in this study are included in the main article and its Supplementary Information files.

## Acknowledgements

The authors would like to thank Dr. Barry Stipe for helpful discussions. This research was funded, in part, by the Data Storage Systems Center at Carnegie Mellon University and all its industrial sponsors and by the Kavcic–Moura Fund at Carnegie Mellon University. The authors acknowledge the use of the Materials Characterization Facility at Carnegie Mellon University supported by Grant No. MCF-677785.

## Author contributions

C.X., D.E.L., J.-G.Z., B.V. contributed to the idea of the study and designed, evaluated, and discussed the experiments. C.X. performed all experimental work. C.X. and J.-G.Z. wrote the paper. All authors reviewed the manuscript and approved the final version.

**Competing Interests**: The authors declare no competing interests.

# Figures

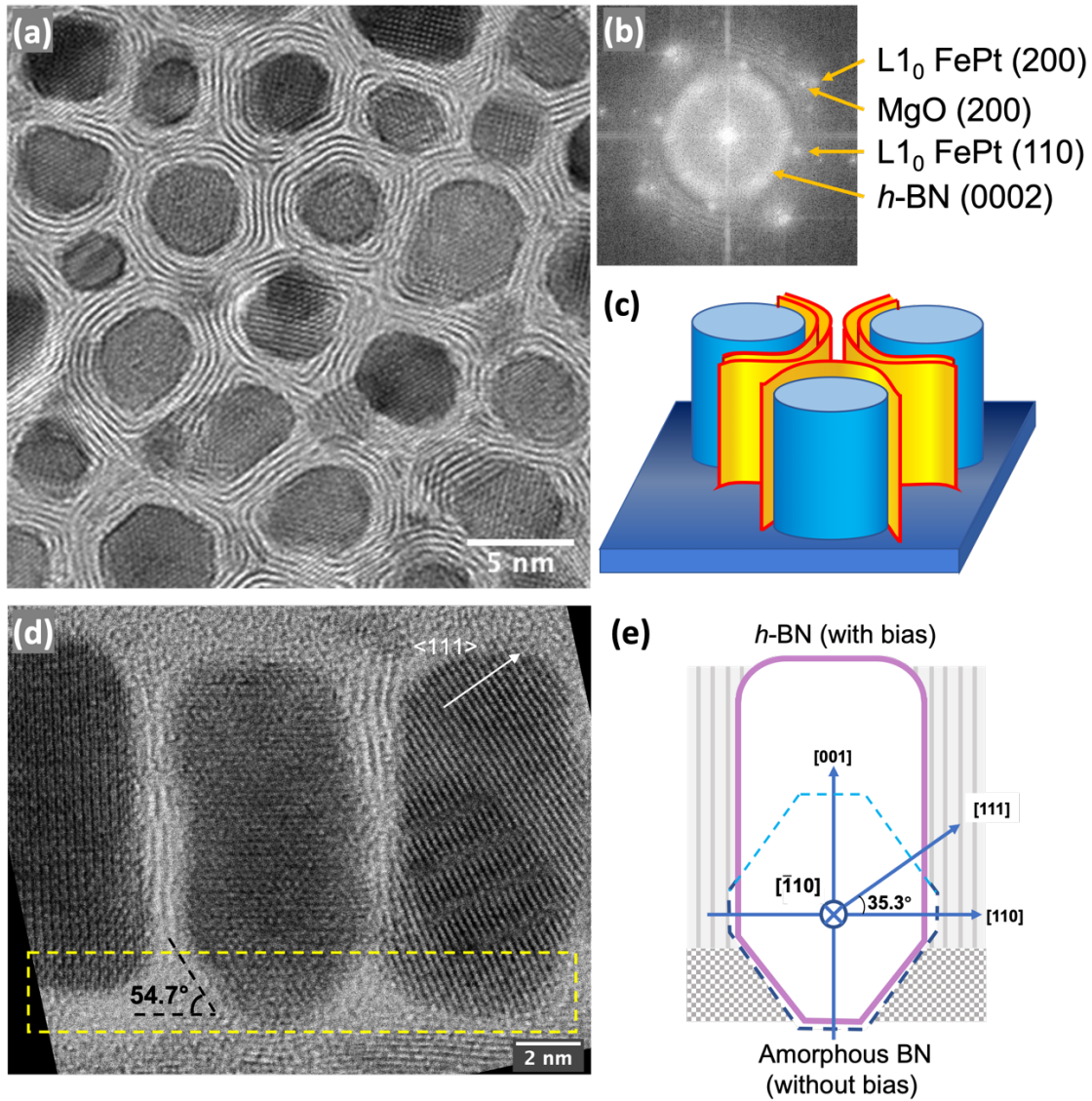

**Figure 1.** TEM micrograph analysis of the FePt/*h*-BN columnar core-shell structure. (a) Plan view HRTEM of Sample-1 and (b) FFT pattern. (c) The schematic of the core-shell nanostructure. (d) STEM-BF cross-sectional image of Sample-2 (lower volume fraction of BN). (e) The schematic demonstrates the viewing axis $[\bar{1}10]$, the grain shape, and the structure of the BN grain boundary materials. The blue dashed line delineates the Wulff shape of $L1_0$ FePt grains at a high temperature viewed from the $[\bar{1}10]$ direction.

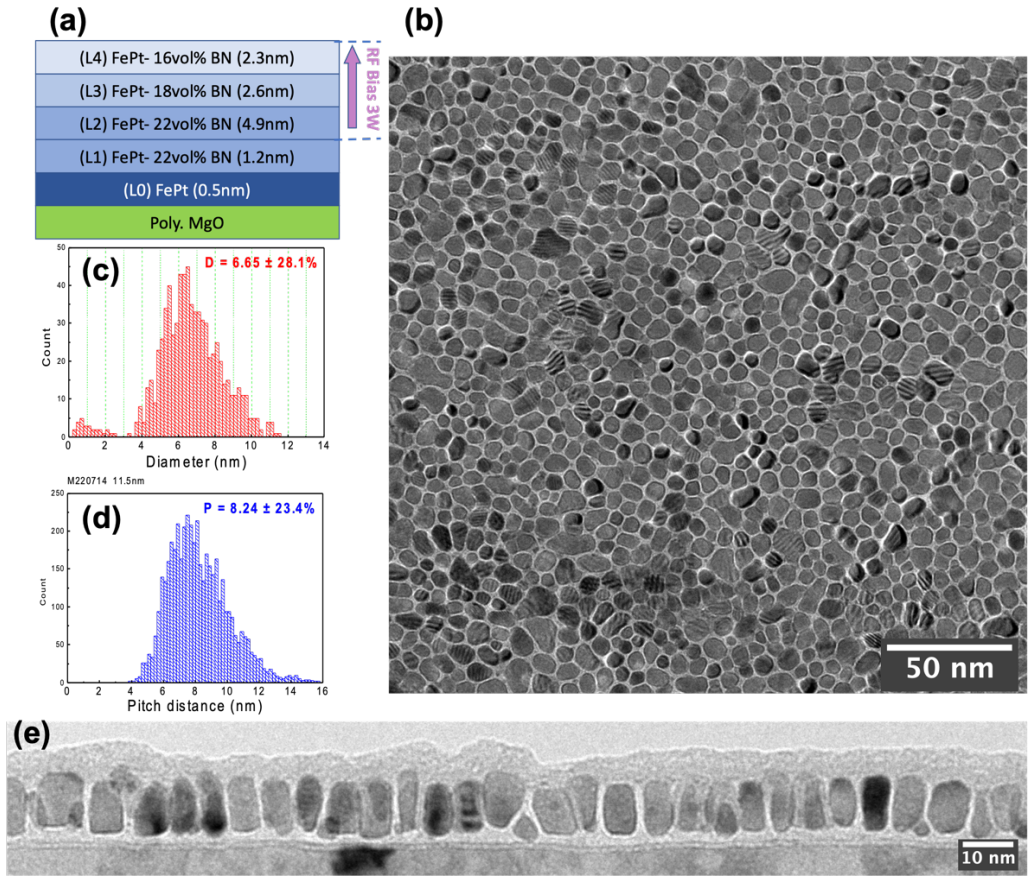

**Figure 2.** TEM analysis of Sample-2 (FePt- 19vol% BN, $t$ = 11.5nm). (a) Diagram of the detailed film stack. (b) BF-TEM plane-view image and (e) cross-sectional image. (c) Grain size distribution. (d) Grain center-to-center pitch distance distribution.

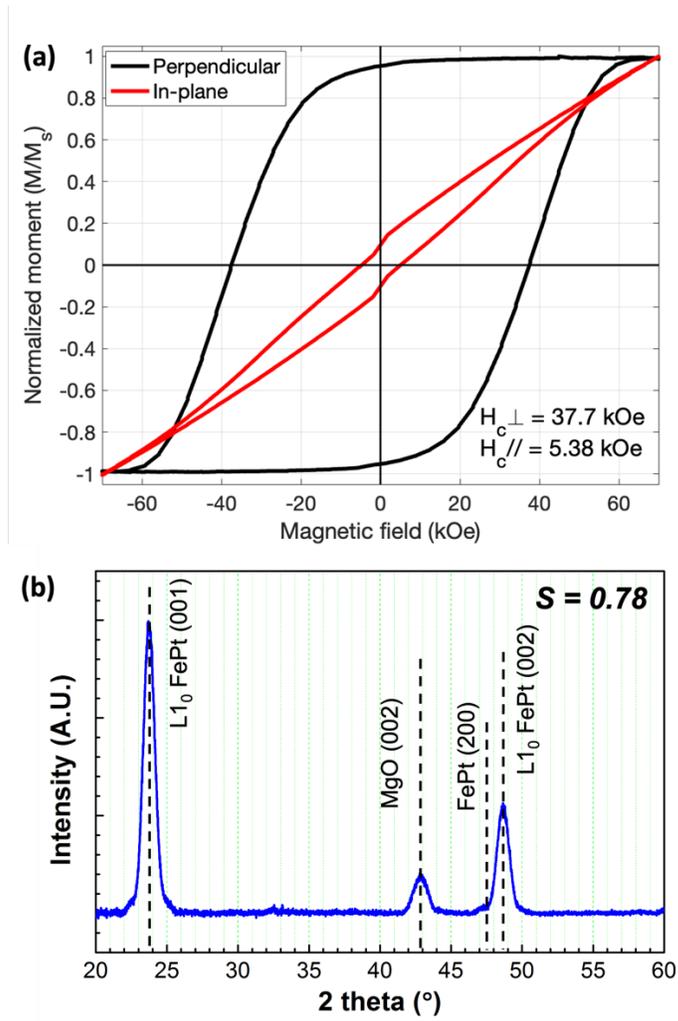

**Figure 3.** (a) Magnetic hysteresis loop and (b) θ–2θ XRD pattern of Sample-2 (FePt- 19vol% BN, $t = 11.5$ nm).

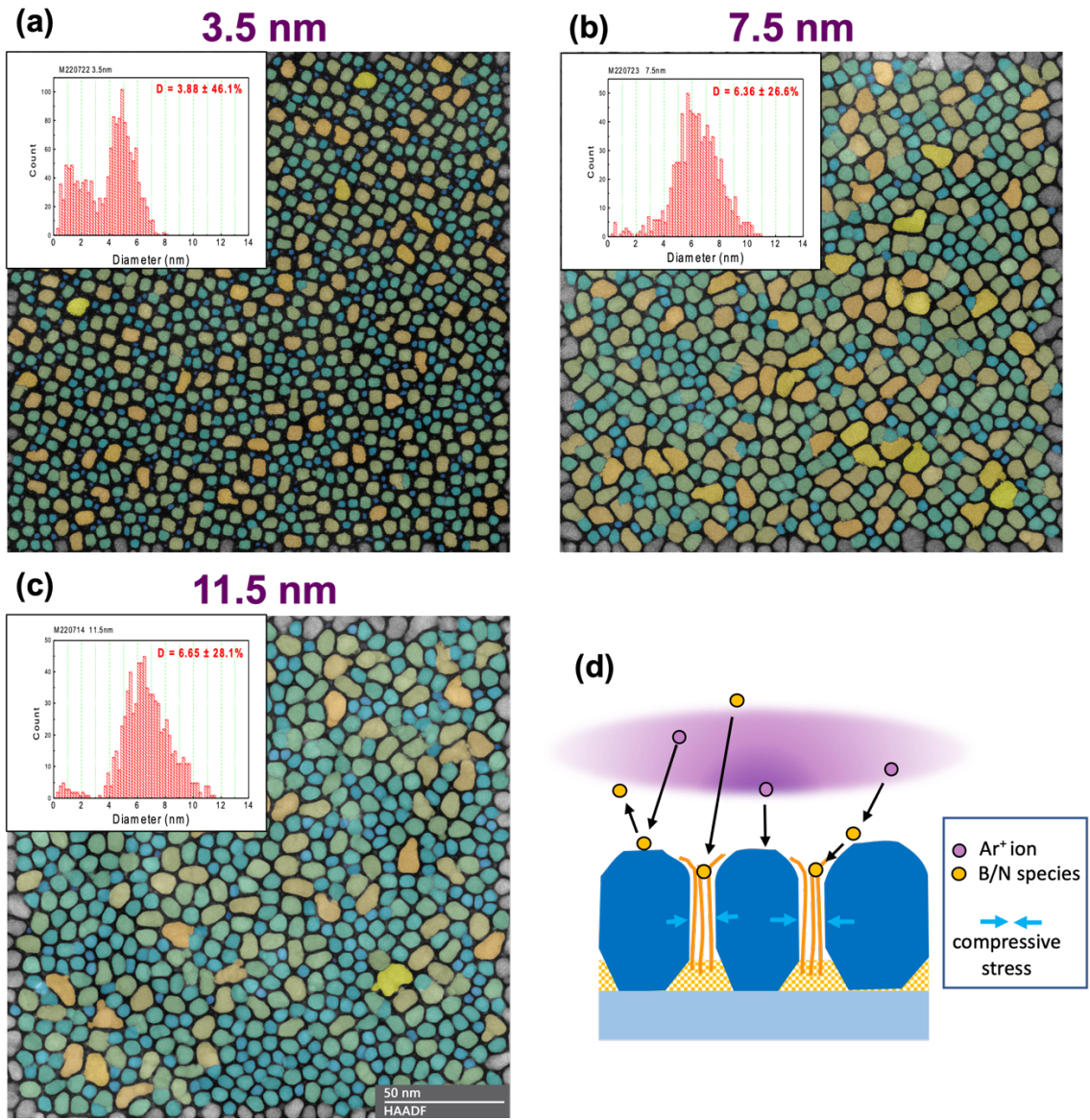

**Figure 4.** Growth process analysis of the FePt/*h*-BN granular film. The false-color STEM-HAADF plane-view images of (a) Sample-2α (*t* = 3.5 nm), (b) Sample-2β (*t* = 7.5 nm), and (c) Sample-2 (*t* = 11.5 nm) share the 50-nm scale bar. Bigger grains are colored yellow and smaller ones are colored cyan. (d) The diagram shows the hypothesized effect of RF bias: resputter effect for kicking off weakly bound atoms with *h*-BN monolayers as the surviving phase.

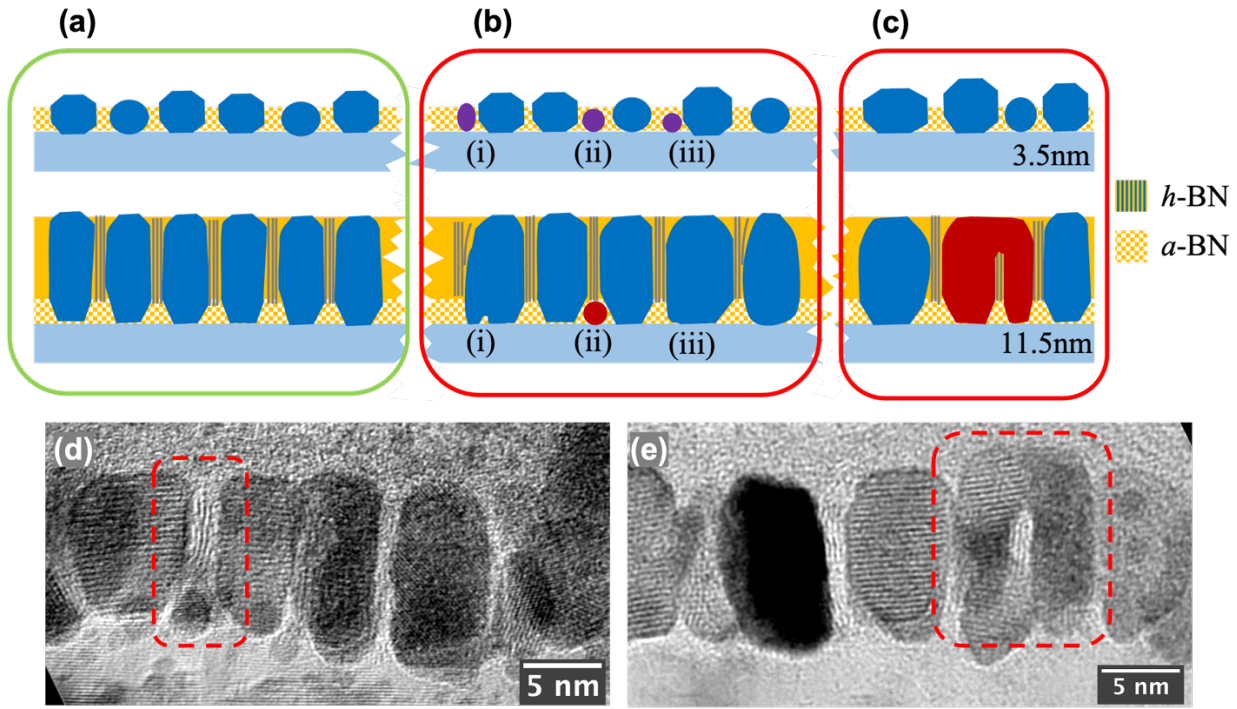

**Figure 5.** Characterization of L1$_0$ FePt granular film with boron nitride grain boundaries. (a) Illustration of columnar growth of the majority of FePt grains: In the FePt/*a*-BN layer, FePt grains exhibit facet growth, with their lateral size increasing with film thickness. With RF-bias turned on, *h*-BN starts to form inside grain boundaries, facilitating the columnar growth of FePt grains. (b) Illustrations of three types of defect growth: i) a small grain closely placed to a large one gradually grow together to form a larger grain; ii) A small FePt grain shadowed by two adjacent FePt grains allowing the *h*-BN grain boundary to grow on top; iii) two small grains coalesce in the early stage of film growth, resulting in a much larger size grain. (c) Lateral top-connection of adjacent FePt grains shielding the growth of *h*-BN grain boundary. (d) Cross-sectional HRTEM image showing clear evidence for the defective growth illustrated in (b); (e) Image showing the evidence for the top-connected adjacent FePt grains illustrated in (c).

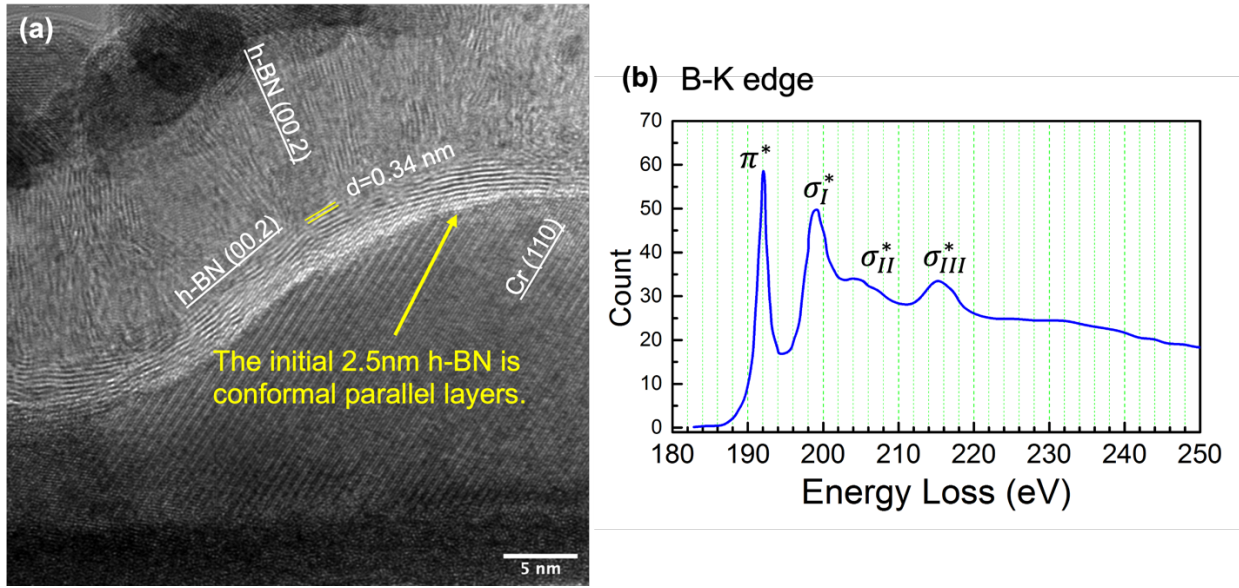

**Figure 6.** (a) Cross-sectional HRTEM image of the 13-nm *h*-BN blanket film on the curved Cr surface. The *h*-BN was RF-sputtered at 650°C, with a 4W RF substrate bias. (b) EEL spectrum acquired on *h*-BN regions, showing the B-K edge with one $\pi^*$ peak (191.8 eV) and three $\sigma^*$ peaks (198.8 eV, 204.2 eV, 215.2 eV).

# Supplementary Information

# Bias sputtering of granular $L1_0$-FePt film with hexagonal boron nitride grain boundaries

Chengchao Xu[1, 2, *], B.S.D.Ch.S Varaprasad[1, 2], David E. Laughlin[1, 2, 3], and Jian-Gang (Jimmy) Zhu[1, 2, 3].

1) Data Storage Systems Center, Carnegie Mellon University, Pittsburgh, Pennsylvania 15213, USA

2) Electrical and Computer Engineering Department, Carnegie Mellon University, Pittsburgh, Pennsylvania 15213, USA

3) Materials Science and Engineering Department, Carnegie Mellon University, Pittsburgh, Pennsylvania 15213, USA

* Correspondence and requests for materials should be addressed to C.X. (email: chengchx@andrew.cmu.edu)

## 1. High-resolution TEM/STEM images of Sample-2

Figure S1(a) shows the HRTEM plane-view image of Sample-2 (FePt-19vol% BN, $t$ = 11.5 nm). Similar to the FFT pattern in Fig. 1(b) for Sample-1 (FePt-38vol% BN, $t$ = 7.5 nm), its FFT pattern also reveals the {110} super-lattice spots of the $L1_0$ FePt ($d$ = 0.271 nm) and a ring corresponding to the $h$-BN (0002) basal planes ($d$ = 0.333 nm). It not only indicates the good chemical ordering and [002] texture of FePt grains but also shows that the conformal fringes in the grain boundary regions come from the $h$-BN nanosheets. In Sample-2 with a relatively low BN volume fraction, the average grain boundary width is ~1.5nm, which can accommodate 3-4 layers of $h$-BN. For the cross-sectional view, the STEM-BF technique was adopted because it prevented the contrast delocalization artifact of HRTEM imaging, as shown in Figure S1(a) in which the bright "shadows" of certain FePt grains displaced from the original locations. Figure 1(d) was cropped from the region framed by the yellow dashed line in Figure S1(b).

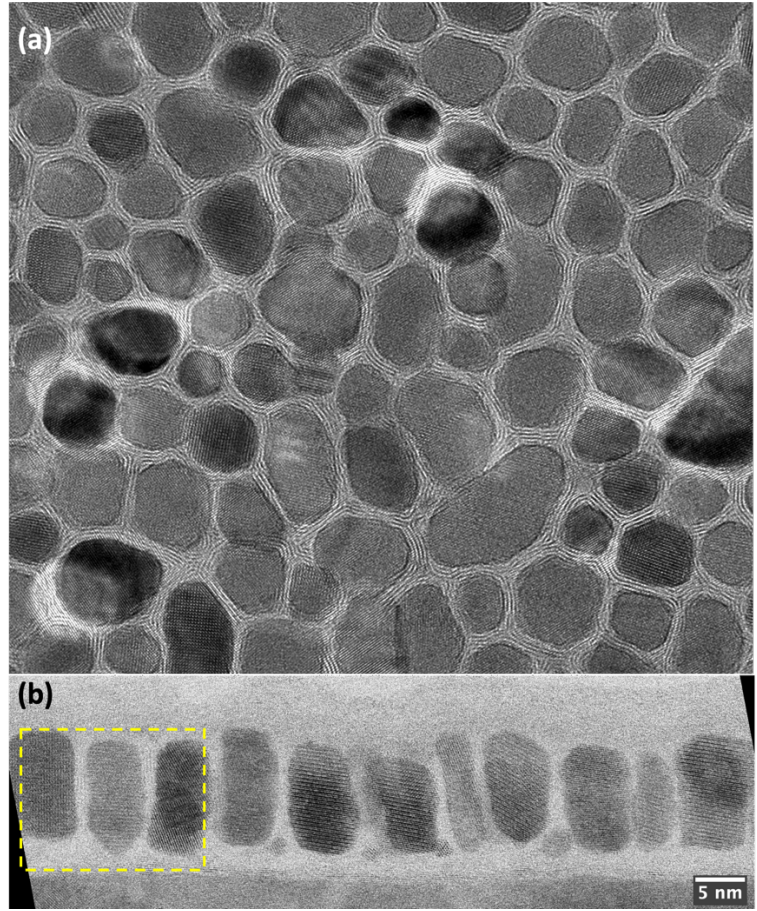

Figure S1. (a) Plane-view HRTEM image. (b)Cross-sectional STEM-BF image of Sample-2 ($t$ = 11.5 nm)

## 2. Growth process analysis of FePt/BN films

To better understand the growth process, Sample-2α ($t$ = 3.5 nm) and Sample-2β ($t$ = 7.5 nm) were fabricated using the same film stack design and deposition conditions as Sample-2 (film stack shown in Fig. 2(a)), except for the thickness. They are intended to represent the three stages of the film growth for Sample-2. The magnetic hysteresis loops and XRD patterns of these samples with varying thicknesses are shown in Figure S2. The high-resolution TEM images of Sample-2α and Sample-2 are presented in Figure S3. As mentioned before, Sample-2α was designed to illustrate the nucleation stage of FePt with $a$-BN grain boundaries deposited without RF bias. In its perpendicular hysteresis loop (Fig. S2(a)), there is a notable soft kink near zero fields, which should be the result of the smaller FePt nuclei (as shown in Fig. S3(a)). Due to the low coordination environment for the atoms on grain surfaces, they experience weaker exchange coupling. Therefore, the smaller FePt grains with higher surface-to-volume ratios normally show lower anisotropy fields or coercivities, contributing to the switching field variance or the Curie Temperature variance ($\sigma T_C$) for HAMR applications. As the film grows thicker, Sample-2β ($t$ = 7.5nm) exhibits a significant improvement of perpendicular coercivity ($H_{C\perp}$) and loop squareness (Fig. S2(b)). This agrees with the change of microstructures (Fig. 4) that the grain size rose dramatically, and the bimodal size distribution disappeared. Only ~8% of the smaller nuclei (D < 3nm) remained during the evolution from Sample-2α to Sample-2β, since they were either shadowed by the adjacent larger grains or coalesced with others to form larger ones. It also contributes to the disappearance of the soft kinks near zero fields. In Sample-2β's $\theta - 2\theta$ XRD pattern (Fig. S2(d)), the peak widths (FWHM) of L1$_0$-FePt (001) and (002) peaks were decreased, and the intensity ratio $I_{(001)}/I_{(002)}$ was higher compared with Sample-2α. Then, during the growth from 7.5nm to 11.5nm, the microstructure and magnetic properties stabilize and change little.

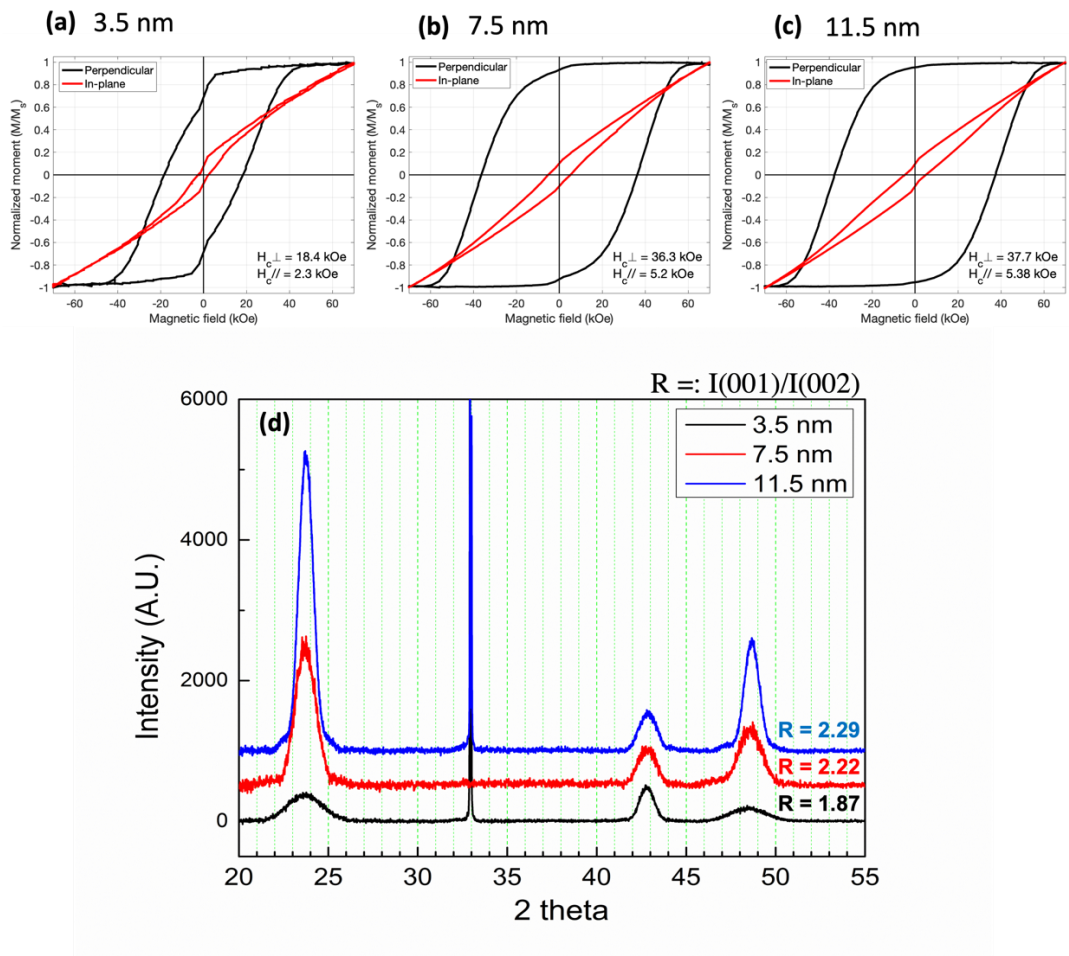

Figure S2. Properties of FePt/BN films with different thicknesses to study the growth processes. (a)-(c) Magnetic hysteresis loops for (a) Sample-2α ($t$ = 3.5 nm), (b) Sample-2β ($t$ = 7.5 nm), (c) Sample-2 ($t$ = 11.5 nm). (d) XRD patterns of the three samples above ($t$ = 11.5, 7.5, 3.5nm from top to bottom).

FePt grains with the desired [002] texture and lateral (projected) grain edges parallel with <110> orientations were obtained in Sample-2α, as shown in Fig. S3(a). The rectangular shapes of nuclei agree with the Wulff polyhedron of $L1_0$ FePt nanocrystals viewed along the [001] direction. Some of the large nuclei in the cross-sectional image, Fig. S3(b), also exhibit {100} and {111} facets. Bimodal size distribution was developed in the nucleation stage. Figure S3(d) showcases a typical microstructure imperfection, lateral grain connection, which is sparsely located in both Sample-2β and Sample-2. Because of the overall deficit of the *h*-BN grain boundary materials (GBM), lateral grain connections on top of *h*-BN at various heights are possible, eventually forming elongated grains or warm-shaped large grains. Figure S3(e) shows an example of defective growth illustrated in Fig. 5(b).(i) and described in the paper.

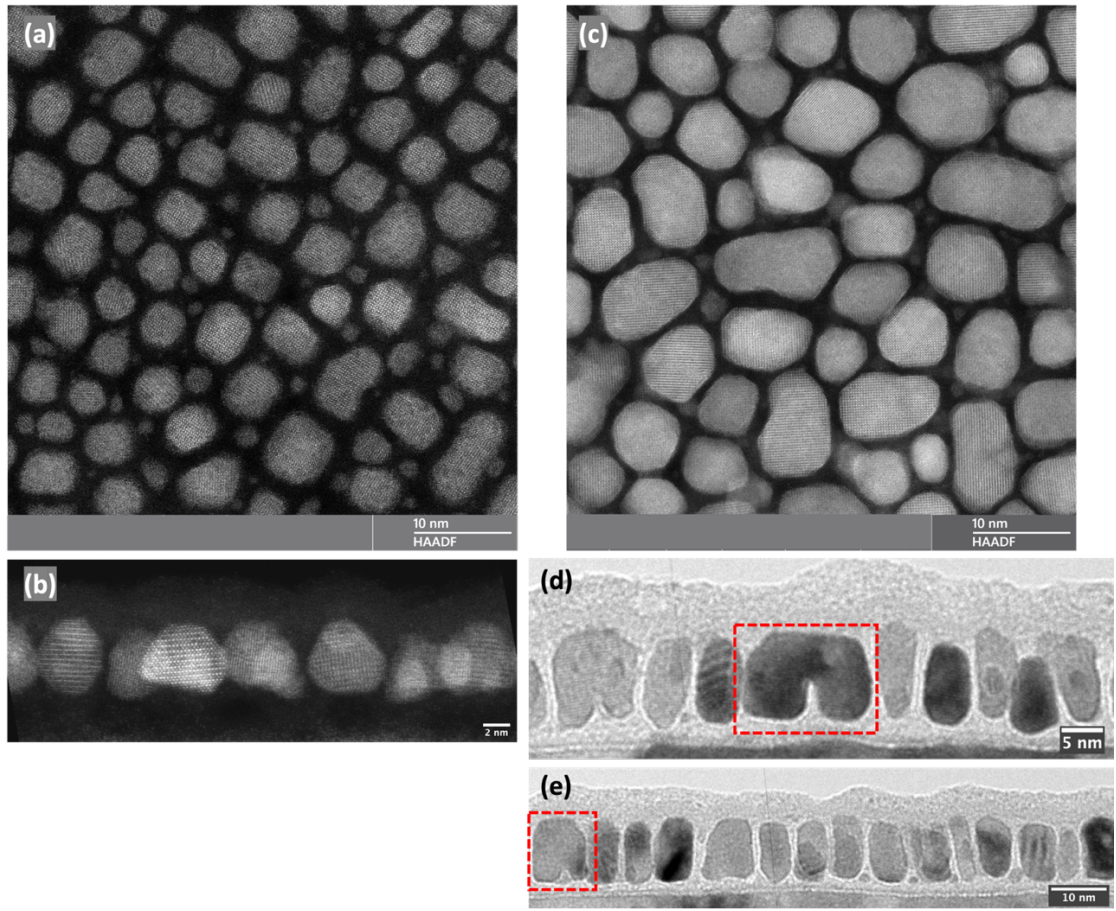

Figure S3. TEM analysis of the microstructure of FePt/BN films with different thicknesses. (Left) The STEM-HADDF plane-view image(a) and cross-sectional image (b) of Sample-2α ($t$ = 3.5 nm). (Right)The STEM-HADDF plane-view image(c) and BF cross-sectional image (d,e) of Sample-2.

## 3. Conditions for *h*-BN formation: high temperature and substrate bias

To Investigate the critical conditions for *h*-BN formation we fabricated three samples with three sets of conditions on FePt (002) layer. The control sample shown in Fig. S4(b) was deposited at 700°C with a substrate bias of 4W to generate perpendicular *h*-BN basal planes, as shown in its FFT pattern. In comparison, removing the bias (Fig. S4(c)) completely amorphized the BN film. Alternatively, when the temperature was lowered to 450°C while the bias was maintained, layered structures were observed in Fig. S4(a), but they were poorly aligned/crystallized. These three results reveal that though both high temperature and substrate bias are necessary conditions for the formation of *h*-BN, substrate bias is the predominant contributor.

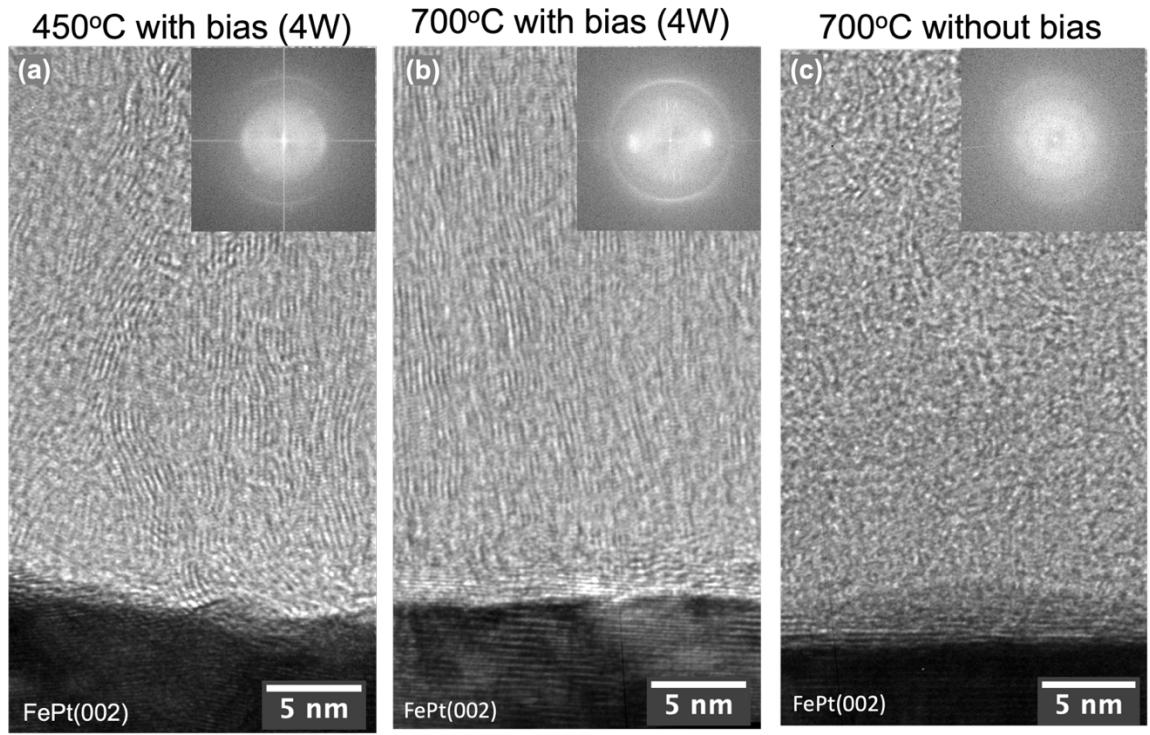

Figure S4. Cross-sectional HRTEM images and corresponding FFT patterns of three BN films deposited on FePt (002) underlayer. (a) 450 °C with RF bias; (b) 700 °C with RF bias; (c) 700 °C without applying bias.